\documentclass[twocolumn,amsmath,amssymb,physrev,superscriptaddress]{revtex4-1}
\pdfoutput=1
\usepackage{graphicx}
\usepackage{color}
\usepackage{float}
\usepackage{dcolumn}
\usepackage{amsmath}
\usepackage{bm}
\usepackage[utf8]{inputenc}
\usepackage{dsfont}

\begin{document}
\title{Nonhermitian defect states from lifetime differences}

\begin{abstract}
Nonhermitian systems provide new avenues to create topological defect states. An unresolved general question is how much the formation of these states depends on asymmetric backscattering, be it nonreciprocal as in the nonhermitian skin effect or reciprocal as encountered between the internal states of asymmetric microresonators.
Here, we demonstrate in a concrete, practically accessible setting  of a lossy coupled-resonator optical waveguide that nonhermitian defect states can exist in open optical systems due to lifetime differences, without the need for asymmetric backscattering within or between the individual resonators. We apply our findings to a finite system of coupled circular resonators perturbed by nanoparticles, following the concept of creating an interface by inverting the position of the nanoparticles in half of the chain. We compare a coupled-mode tight-binding approximation to full-wave numerical simulations, showing that spectrally isolated defect states can indeed be implemented in this simple nonhermitian photonic device.
\end{abstract}

\author{Martí Bosch}
\affiliation{Institut f{\"u}r Physik, Technische Universit{\"a}t Ilmenau, D-98693 Ilmenau, Germany}
\affiliation{Department of Physics, Lancaster University, Lancaster LA1 4YB, United Kingdom}
\author{Simon Malzard}
\affiliation{Department of Physics, Lancaster University, Lancaster LA1 4YB, United Kingdom}
\affiliation{Department of Mathematics, Imperial College London, London SW7 2AZ, United Kingdom}
\author{Martina Hentschel}
\affiliation{Institut f{\"u}r Physik, Technische Universit{\"a}t Ilmenau, D-98693 Ilmenau, Germany}
\author{Henning Schomerus}
\affiliation{Department of Physics, Lancaster University, Lancaster LA1 4YB, United Kingdom}

\maketitle

\section{Introduction}
Nonhermitian physics has attracted tremendous interest in the past decade, not least due to the variety of physical systems that can be captured by nonhermitian effective Hamiltonians,
such as in condensed-matter physics \cite{Hatano.1997,Longhi.2009b}, optomechanics \cite{Jing.2017} and photonics \cite{Musslimani.2008, Ruter.2010, Chong.2011, Peng.2014}. Of these, settings in optics and photonics have been recognized as particularly suitable platforms due to the parallels between the Schr{\"o}dinger equation and Maxwell's theory of light \cite{Longhi.2009}.
Nonhermiticity can be introduced in photonic systems by two distinct routes, where the first considers gain and loss, while the second considers asymmetric and potentially nonreciprocal coupling mechanisms (asymmetric coupling, AC).
These, in addition to the intrinsic openness of optical systems, lead to the occurrence of nonhermitian features with a wide range of applications, especially through the manipulation of exceptional points \cite{Wiersig.2014, Peng.2016, Chen.2017, Kullig.2018}, as they occur generically, e.g., in PT-symmetric systems \cite{ElGanainy.2018}.

In recent years, nonhermitian physics has been further enhanced by the recognition of topological effects, which are based on the interplay of a wide range of symmetries going beyond the PT case, and manifest themselves in a  variety of bulk and boundary phenomena, including novel interface and defect states \cite{Poli.2015, Malzard.2015} and  bulk and boundary Fermi arcs \cite{Ni.2018,Zhou.2018,Malzard.2018, Carlstrom.2018}.
In particular, nonhermitian defect states equipped with a topological mode selection mechanism \cite{Schomerus.2013,Poli.2015} have already been exploited for the design of lasers on a variety of platforms \cite{StJean.2017,Zhao.2018,Yao.2018b,Ota.2018,Whittaker.2019}.
In some cases, such defect states can still be characterized by mappings to the hermitian topological setting, and thereby remain associated with bulk and boundary invariants that conform with the bulk-boundary principle \cite{Gong.2018,Kawabata.2018a}. For these simple settings, a paradigmatic example is a Su-Schrieffer-Heeger (SSH) chain with a complex potential as realized by gain, loss, or other dissipative mechanisms \cite{Rudner.2009,Schomerus.2013}.
Remarkably, however, it has also been established that new classes of topologically robust interface and defect states can emerge in nonhermitian settings that would be topologically trivial in their hermitian (closed system) limit.
Based on the study of a variety of specific systems,
two distinct, well defined mechanism have so far been identified.

The first mechanism is the nonhermitian skin effect (SE), which is intimately related to nonreciprocal AC and can be understood, alternatively, from the ensuing nonconserved probability flux, the exponential distortion of the probability weights in right and left eigenstates when compared to the symmetric coupling case, and the proximity to high-order exceptional points when the coupling asymmetry is taken to the extreme. Subject to this SE mechanism, systems are highly sensitive to the boundary conditions, also entailing that the bulk-boundary principle has to be revisited \cite{Kunst.2018}. Paradigmatic examples in this first class of essentially nonhermitian topological systems are the Hatano-Nelson model \cite{Hatano.1996}, as well as variants of the SSH models with nonreciprocal AC \cite{Yao.2018, Yin.2018}.

In the second mechanism, defect states appear in reciprocal gain-loss settings at a sufficiently strong level of nonhermiticity via exceptional points (EP), signifying that scattering solutions turn into normalizable solutions. Even though the scattering solutions pertain to the band structure, the bulk-boundary principle in its original form is again violated, as the band structure itself does not drastically change at the EP. In this case topological protection is understood in terms of the robustness of the EPs in parameter space, while a general theory of bulk and boundary invariants has not yet been developed. The paradigmatic
candidate example of this second class of essentially nonhermitian topological systems is a reciprocal lossy resonator chain (a lossy coupled-resonator optical waveguide, CROW), for which, the EP mechanism  has only been described assuming reciprocity-conserving AC between internal resonator modes \cite{Malzard.2015, Lang.2018}.

So far, both the SE and EP mechanism have been mainly explored in coupled-mode tight-binding models, and have not yet been realized in photonic experiments.
On paper, the most promising route to nonreciprocal AC follows the steps of hermitian photonic topological insulators \cite{Hafezi.2013}, which are based on evanescently coupled ring resonators with effectively decoupled clockwise (CW) and counterclockwise (CCW) propagation sectors. The time-reversal operation maps both sectors onto each other, but within each sector time-reversal symmetry is effectively broken. Again focussing on each sector, nonreciprocal AC can then in principle be induced by lossy elements placed into auxiliary rings  \cite{Longhi.2015}, even though this has not yet been demonstrated in practice.
Reciprocal AC, on the other hand, follows generically in asymmetric, open resonators \cite{Wiersig.2008}, and outside the topological setting has been observed experimentally for a wide range of realistic individual resonator shapes \cite{Cao.2015}.

For the design of experiments and applications, this leaves two important open questions. Firstly, from a more practical perspective,  can such realistic resonator shapes induce defect states via the EP mechanism if placed into an appropriate resonator chain? Secondly, from a more fundamental point of view, is reciprocal AC within these resonators a key ingredient, or can the same effects also achieved in simpler symmetric shapes, hence based on the more conventional, manifestly reciprocal standing-wave combinations of the CW and CCW waves?

\begin{figure}[t]
\includegraphics[width=\columnwidth]{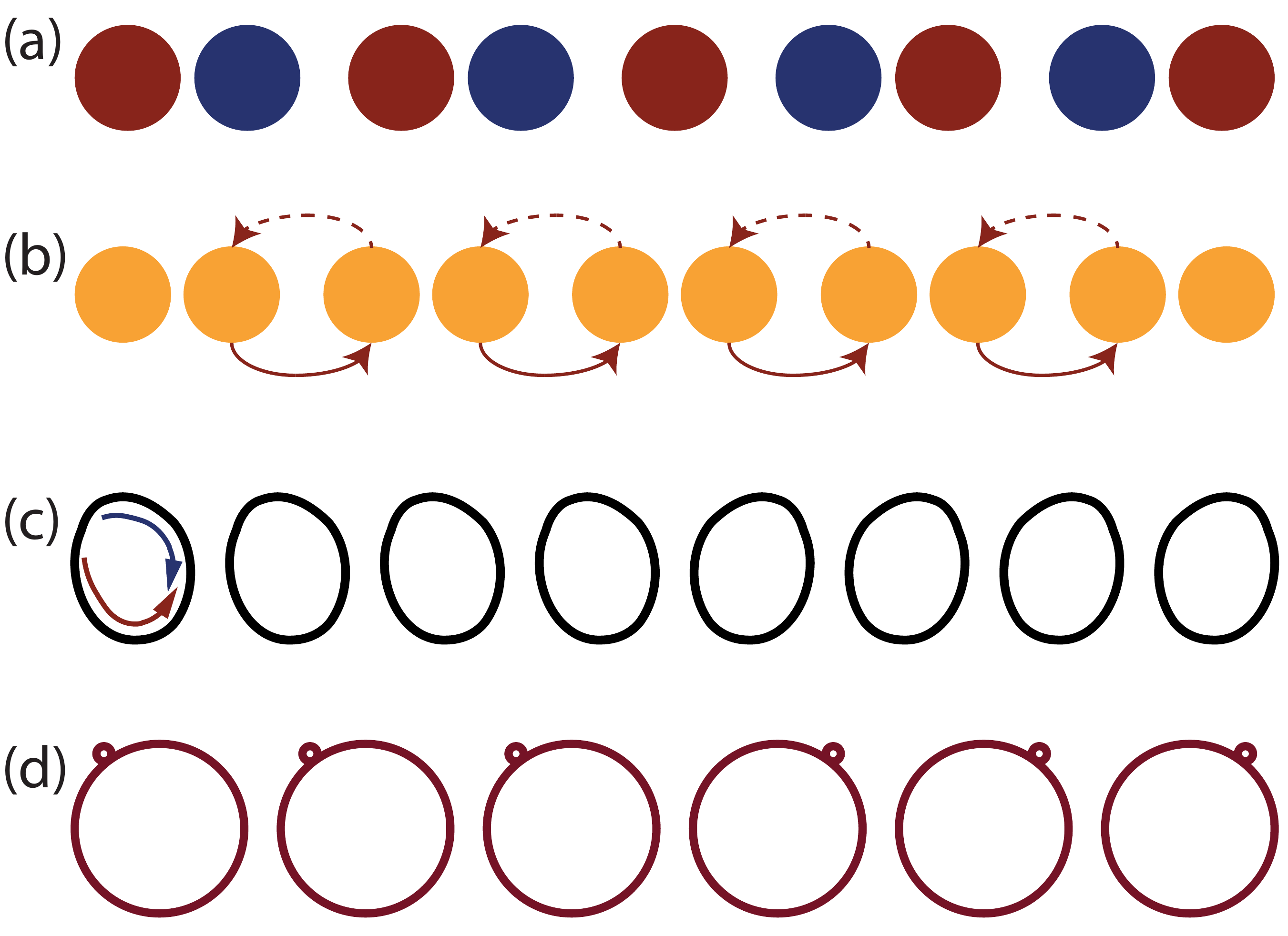}
\caption{Design concepts of nonhermitian systems with defect states. (a) Systems with distributed gain (red) and loss (blue), as, e.g., realized in topological mode selection \cite{Poli.2015}. (b,c) Systems with asymmetric  coupling (indicated by arrows), either external and nonreciprocal (b) as exploited in the nonhermitian skin effect \cite{Yao.2018} or internal and reciprocal (c) as in open lossy asymmetric resonator chains \cite{Malzard.2015}.  (d)
In this paper, we show that simpler systems with lifetime differences in otherwise degenerate modes can display similar characteristics as systems with reciprocal asymmetric coupling. These lifetime difference can, e.g., be obtained by perturbing symmetric resonator shapes by nanoparticles.}
\label{idea}
\end{figure}

In this paper, we demonstrate that even simple resonator chains, based on symmetric individual resonator shapes without internal or external AC, can indeed lead to the desired defect states. This requires neither the incorporation of material gain or loss as illustrated for the example of topological mode selection in Fig.~\ref{idea}(a), nor the asymmetric backscattering employed in the SE and EP mechanisms as illustrated  in Fig.~\ref{idea}(b,c). Rather, we find that the interface states arise from lifetime differences due to the leakage of the supported standing-wave modes, which can be induced by very simple means as depicted in Fig.~\ref{idea}(d).

The general design of the system, which is based on the same generic setting as  the paradigmatic model in Ref. \cite{Malzard.2015} and can be realized as a lossy CROW \cite{Schomerus.2014}, is presented in Sec. \ref{sec:sys}.
Using first the tight-binding approximation (Sec. \ref{sec:res1}), we show that defect states can arise even in absence of AC if the system is perturbed in a way such that the lifetimes of an eigenmode pair split.
Moving towards practical settings (Sec. \ref{sec:res2}),
we then identify a suitable resonator geometry, obtained by perturbing a circular resonator with a single nanoparticle, a setting where internal AC is known to be absent \cite{Wiersig.2011b}. Furthermore, we compare the results of the tight-binding approach with numerical simulations of the realistic system, implemented by incorporating the individual resonators into a chain with two different positions of the nanoparticles on two sides of an interface. Our conclusions are collected in Sec. \ref{sec:con}.

\section{Background and methods\label{sec:sys}}
The systems studied in this work are lossy CROWs composed of planar, almost circular dielectric microresonators.
We follow the route of most applications, where such microresonators are grown with a low aspect ratio, allowing them to be treated as two-dimensional structures with an effective refractive index $n(x,y)$. In this case Maxwell's equations can be reduced to the scalar wave equation \cite{Jackson.1999}
\begin{equation}
\label{eq:1}
- \nabla^2 \psi = n^2(x,y) \frac{\omega^2}{c^2}\psi ,
\end{equation}
where $\omega = c k$ is the frequency, $c$ the speed of light in vacuum and $k$ the wavenumber. For systems with a piecewise constant refractive index, Eq. \eqref{eq:1} is valid for both transversal electric (TE) and transversal magnetic (TM) polarization, for which the electric or magnetic field vector lies in the cavity plane; the only difference are the matching conditions at the boundaries of the regions of constant refractive index.
The nonhermiticity of the system arises from its openness, which can be considered by applying Sommerfeld outgoing wave conditions at infinity \cite{Schot.1992}. The wave equation for the TM modes was solved numerically using the finite element method software Comsol Multiphysics\textsuperscript{\textregistered} 5.3, wave optics module \cite{COMSOL}, where perfectly matched layers were used to simulate the openness of the system \cite{Berenger.1994}.

In order to describe these systems more conceptually, we use a coupled-mode tight-binding Hamiltonian, which allows to approximate the complex eigenfrequencies of the system in the relevant spectral range. This is achieved by extending the two-mode approximation used to describe the main properties of almost-circular single resonators \cite{Wiersig.2008} to a chain of coupled resonators \cite{Schomerus.2014}.

 The two-mode approximation focusses on a pair of whispering gallery modes (WGMs)  with  CW and  CCW orientation, which are described by an effective nonhermitian two-by-two Hamiltonian
\begin{equation}
\label{eqn:a} H = \begin{pmatrix}
\Omega_0  & 0\\
0 & \Omega_0\\
\end{pmatrix}+\begin{pmatrix}
\Omega_p  & A\\
B & \Omega_p\\
\end{pmatrix} = \begin{pmatrix}
\bar\Omega  & A\\
B & \bar\Omega\\
\end{pmatrix}
\end{equation}
whose complex eigenvalues correspond to the eigenfrequencies of the mode pair.
In the perfectly circular system, these eigenfrequencies are identical for both modes, and denoted as $\Omega_0$. To account for a small deformation or perturbation of this situation, we include a perturbation term to the Hamiltonian, which both shifts the frequency of both modes by a small, still identical amount $\Omega_p$, but also introduce backscattering coefficients $A$ and $B$. These backscattering coefficients can in principle be nonidentical, corresponding to AC. Importantly, the coefficients appearing here are constrained by reciprocity as well as symmetries of the resonator shape, as we will explain in detail in Sec.~\ref{sec:tb} further below, and forms the basis of our main results.

This model is sufficient to describe the effects of backscattering and  openness for the mode pair in a single resonator. Furthermore, under the assumption that the WGM mode pair only couples to analogous mode pairs in the neighboring resonators, this description can be extended to a system of $N$ coupled resonators. This gives an $N \times N$ block matrix $\cal H$ with blocks $H_n$ in the diagonal elements and blocks $T$ in the next-to-diagonal elements, where
\begin{equation} H_n =\begin{pmatrix}
\bar\Omega  & A_n\\
B_n & \bar\Omega \\
\end{pmatrix}, \qquad T =\begin{pmatrix}
0  & W\\
W & 0\\
\end{pmatrix}.
\end{equation}
Here we assume that CCW (CW) modes couple only to the CW (CCW) modes in the neighboring resonators, where the coupling constant $W$ is real and equal for all modes, as can be realized in the weak-coupling regime of evanescently coupled resonators \cite{Schomerus.2014}.

The corresponding wave equation is given by
\begin{equation}
\omega \psi_n = H_n \psi_n + T (\psi_{n+1} + \psi_{n-1}).
\end{equation}
We note that the system obeys a nonhermitian chiral sublattice symmetry
\begin{equation}
\sigma_z(\mathcal{H}-\bar\Omega\openone)\sigma_z= \bar\Omega\openone-\mathcal{H}
\label{eq:chiral}
\end{equation}
  with the Pauli matrix $\sigma_z$,
which dictates that the complex frequency spectrum is inversion-symmetric about the frequency $\bar\Omega$.

For an infinite one-dimensional periodic chain, $H_n\equiv H$ for all $n\in\mathbb{Z}$, the solutions of the system are obtained from a superposition of Bloch waves $\psi_n = \Psi e^{ikn}$, fulfilling the  equation
\begin{equation}
\omega(k) \Psi = (H+2\cos k T) \Psi.
\end{equation}
The dispersion relation is given by $\omega(k)=\bar\Omega\pm\sqrt{(A+2W\cos k)(B+2W\cos k)}$, displaying a symmetry about $\bar\Omega$ as dictated by the chiral symmetry \eqref{eq:chiral}.

It is useful to specify this dispersion further for the special case where $A=-B$ are both real, which we will encounter further below, where
\begin{equation}
\label{eq:disp}
\omega(k)=\bar\Omega\pm\sqrt{A^2-4W^2\cos^2 k}.
\end{equation}
For $|A|<2W$ (and $A$ still real), the square root then gives rise to a gapless dispersion relation, where two branches aligned along the real axis are joined up with two branches aligned along the imaginary axis. For $|A|\geq 2W$, on the other hand, we obtain a gapped dispersion with two separate branches aligned along the imaginary axis.

The specific variant of this system analyzed in the present paper consists of a chain of perturbed resonators with an interface, created by inverting the orientation of the resonators in half of the system. The interface can be implemented in the Hamiltonian by using $H_n\equiv H$ for the diagonal elements with $n > N/2$ and $H_n\equiv H^T$ for the diagonal elements with $n \leq N/2$. Overall, the system is then described by a $2N\times2N$ matrix, with its eigenvalues corresponding to the eigenfrequencies of the system arising for a chosen pair of WGMs.

Assuming real $A$ and $B$ to observe an effective PT symmetry, but without considering geometric symmetry constraints on the couplings, it is known that defect states can form by the EP mechanism at sufficiently strong nonhermiticity $(A-B)/W$, which generically embodies reciprocal AC. The symmetry protection of the states arises already if the general Hamiltonian in equation \eqref{eqn:a} displays PT and CT symmetry \cite{Malzard.2015}. By revisiting these conditions in detail, we identify a simplified situation without AC that achieves a formally equivalent effect  (Sec. \ref{sec:res1}), and then show how this situation can be realized in practice (Sec. \ref{sec:res2}).

\section{Lifetime backscattering in the tight-binding approximation\label{sec:res1}}

\subsection{Symmetry constraints\label{sec:tb}}
To understand the emergence and role of reciprocal AC, we first investigate a number of relevant symmetry constraints in the tight-binding model. We both adopt the WGM basis of CCW modes $|+\rangle$ and CW modes $|-\rangle$ (with the symbols denoting the mathematical orientation of the propagation direction in these modes, which have a general angular mode dependence $e^{\pm i m \varphi}$ with azimuthal mode-pair index $m$), as well as their properly normalized standing-wave (SW) counterparts $|c\rangle=\frac{1}{\sqrt{2}}(|+\rangle +|-\rangle)$ (essentially, a cosine wave in the angular dependence) and $|s\rangle=\frac{1}{\sqrt{2}i}(|+\rangle -|-\rangle)$ (essentially, a sine wave).

As the first constraint we consider reciprocity, which arises due to the scalar wave nature of the underlying wave equation \eqref{eq:1}. This constraint is most easily implemented in the standing-wave basis, where the effective Hamiltonian must be symmetric,
\begin{equation}
H^{(SW)} =\begin{pmatrix}
\Omega_{c}  & \Delta \\
\Delta & \Omega_{s} \\
\end{pmatrix}.
\label{eq:hsw}
\end{equation}
The diagonal terms can differ, and are conveniently written as
\begin{align}
\Omega_c&=\bar\Omega+\delta,\\
\Omega_s&=\bar\Omega-\delta,
\end{align}
where
\begin{align}
\bar\Omega=\frac{1}{2}(\Omega_{c}+\Omega_{s}),\\
\delta=\frac{1}{2}(\Omega_{c}-\Omega_{s}).
\end{align}

Translated into the WGM basis we obtain the form
\begin{equation}
H^{(WGM)} =\begin{pmatrix}
\bar\Omega  & A_0\\
B_0 & \bar\Omega \\
\end{pmatrix},
\label{eq:hwgm}
\end{equation}
with the diagonal elements $\bar\Omega$ identical as anticipated in Eq. \eqref{eqn:a},
while the coupling coefficients
\begin{align}
A_0&=\delta-i\Delta,
\nonumber \\
B_0&=\delta+i\Delta
\label{eq:ab}
\end{align}
are general complex numbers, even in the given reciprocal case. This is the sought-after manifestation of reciprocal internal AC.

Symmetric coupling (i.e., absence of AC) occurs when $|A_0|=|B_0|$. Here, $A_0$ and $B_0$ become identical when $\Delta=0$. This situation is readily achieved in resonators with a reflection symmetry, placed suitably to preserve the SW modes $|c\rangle$ and $|s\rangle$. As both modes have a different parity under the reflection, the symmetry of the system prevents their mixing, which directly entails
$\Delta=0$ in the SW basis. Formally, this constraint is born out by the relation $H^{(SW)}=\sigma_z H^{(SW)}\sigma_z$.
Furthermore, this constraint is consistent with the oberservation that the reflection interchanges the WGM modes, meaning that $H^{(WGM)}=\sigma_x H^{(WGM)}\sigma_x$, which again implies $A_0=B_0$.

In order to obtain a nonhermitian system in the absence of reciprocal AC, we require that $A_0=B_0$ is complex. The largest level of nonhermiticity is then achieved when
\begin{equation}
{\rm Re}\,\Omega_c={\rm Re}\,\Omega_s,
\end{equation}
so that
\begin{equation}
A_0=B_0 =\frac{i}{2}{\rm Im}\,(\Omega_c-\Omega_s)\equiv i\,\mathrm{Im}\,\delta
\end{equation}
is purely imaginary. Within the two-mode model, we readily see from Eq. \eqref{eq:ab} that this corresponds to a setting where $\Omega_{c}$ and $\Omega_{s}$ agree in their real parts, hence scatter resonantly at the same real frequency, but differ in their imaginary parts, hence their lifetimes (or, equivalently, display different linewidths). Summarizing these considerations, to achieve strong nonhermiticity in absence of internal reciprocal AC we should therefore aim at symmetric resonator geometries in which a WGM mode pair is split only in lifetime, but not in the real frequency.

\subsection{Nonhermitian defect states}

In order to see whether nonhermitian defect states can arise form these lifetime differences, we make use of one more freedom in the extended two-mode model for the chain, namely, the orientation axis of the resonator's reflection symmetry with respect to the coupling axis. Upon rotation of the symmetry axis by an angle $\beta$, the WGMs transform as $|\pm\rangle \to \exp(\pm im\beta) |\pm\rangle$, so that the effective resonator Hamiltonian takes the more general form
\begin{equation}
H^{(WGM)} =\begin{pmatrix}
\bar\Omega  & i\,(\mathrm{Im}\,\delta) e^{-2im\beta}\\
i\,(\mathrm{Im}\,\delta) e^{2im\beta} & \bar\Omega \\
\end{pmatrix}
\end{equation}
where $\delta$ and $\bar\Omega$ remain defined as in the previous section.

It follows that the offdiagonal elements of the Hamiltonian in (2) remain of the  same magnitude, $|A|=|B|$, so that symmetric backscattering is preserved in any WGM basis. In particular, these offdiagonal elements will be real and opposite,
\begin{equation}
A= -B = (-1)^n\mathrm{Im}\,\delta,
\label{eq:ab1}
\end{equation}
if the following condition is met:
\begin{equation}
2m\beta = \frac{\pi}{2} + n\pi \qquad n \in \mathbb{Z}.
\label{eq:betacond}
\end{equation}

These symmetry considerations demonstrate that it is possible to obtain an effective internal Hamiltonian with the desirable properties $A=-B$ and $A, B \in \mathbb{R}$ by exploiting the lifetime differences of the modes. In principle, this should allow us to meet the conditions for the observation of defect states by the EP mechanism when the nonhermiticity is sufficiently large.

\section{Lifetime backscattering with realistic resonator shapes\label{sec:res2}}
We now explore how these defect states can be realized in realistic systems, where we set out to induce the lifetime-induced backscattering  via small perturbations into almost-circular resonator geometries.  As we will see, this allows us to design a simple optical system with the desired effective Hamiltonian.

\subsection{Single resonator design\label{sec:reson}}

\begin{figure}[t]
\includegraphics[width=.7\columnwidth]{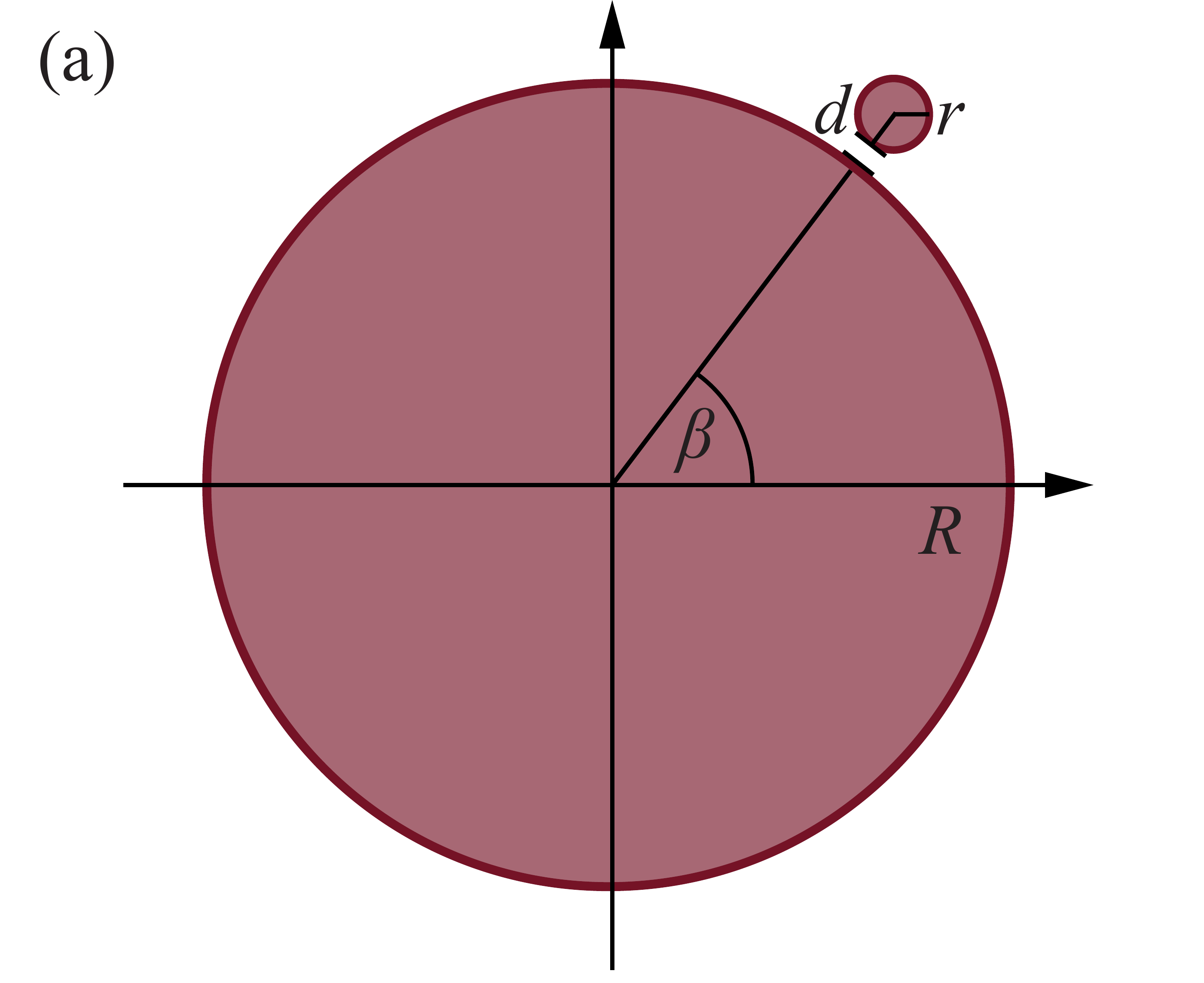}
\includegraphics[width=\columnwidth]{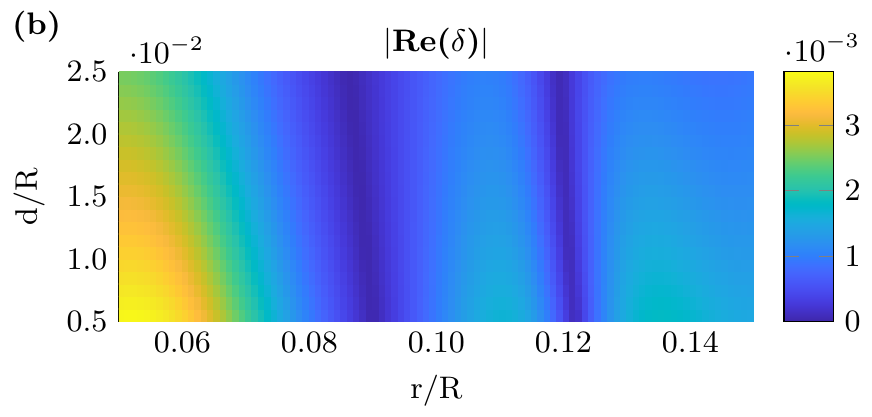}
\includegraphics[width=\columnwidth]{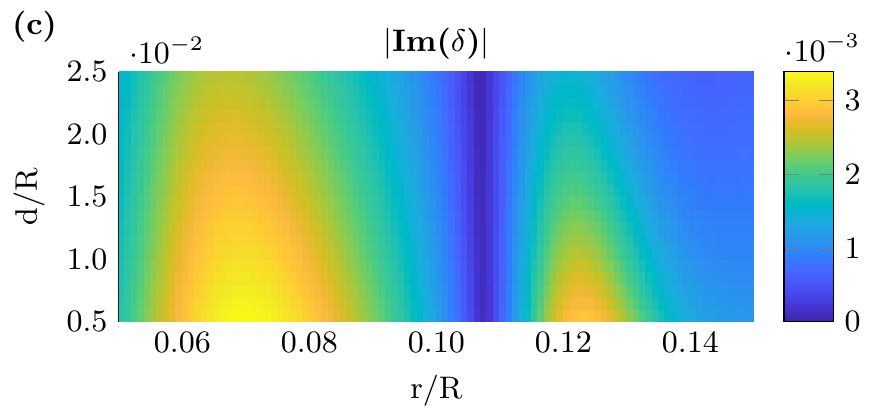}
\caption{
(a) Circular microresonator of radius $R$ and refractive index $n$ with a local perturbation caused by a nanoparticle with the same refractive index as the disk. The position and size of the nanoparticle are described by the effective radius $r/R$, distance $d/R$, and angular position $\beta$.
(b,c)
Complex relative resonance shift $\delta$ for the almost-degenerate WGM mode pair with azimuthal mode number $m = 16$,
 as a function of the relative radius $r/R$ and distance $d/R$. We aim to operate at conditions where $\mathrm{Re}\,\delta=0$, which can be achieved by tuning the single parameter $r/R$ at fixed $d/R$.}
 \label{fig:sweeps}
\end{figure}
In circular resonator geometries, the WGMs in a mode pair $|\pm\rangle$ have degenerate eigenfrequencies $\Omega_0$, which also carry over to their standing-wave combinations  $|c\rangle$ and  $|s\rangle$.
Consider now a local perturbation at the angular position $\phi = \beta$ [cf. Fig.~\ref{fig:sweeps}(a)]. This naturally affects the  even and odd eigenmodes $|c\rangle$ and  $|s\rangle$  differently,
as only  $|c\rangle$ has a large weight at the perturbation position while   $|s\rangle$ has a node. Furthermore, as discussed before, both modes do not hybridize as long as the perturbation respects the symmetry about the reflection axis with angle $\beta$. Applying standard perturbation theory \cite{Wiersig.2008}, the system is described by the SW Hamiltonian \eqref{eq:hsw} with distinct eigenfrequencies $\Omega_c$ and $\Omega_s$and $\Delta=0$.
The real part of the complex frequency splitting $\Omega_c-\Omega_s=2\delta$ corresponds to a relative frequency shift, while the imaginary part corresponds to the lifetime differences of the perturbed modes.
According to the considerations in the previous section, our goal is now to find a perturbation for which $\delta$ is purely imaginary---the lifetimes of the modes are different yet their frequencies are equal.

\begin{figure}[t]
  \includegraphics[width=0.47\columnwidth]{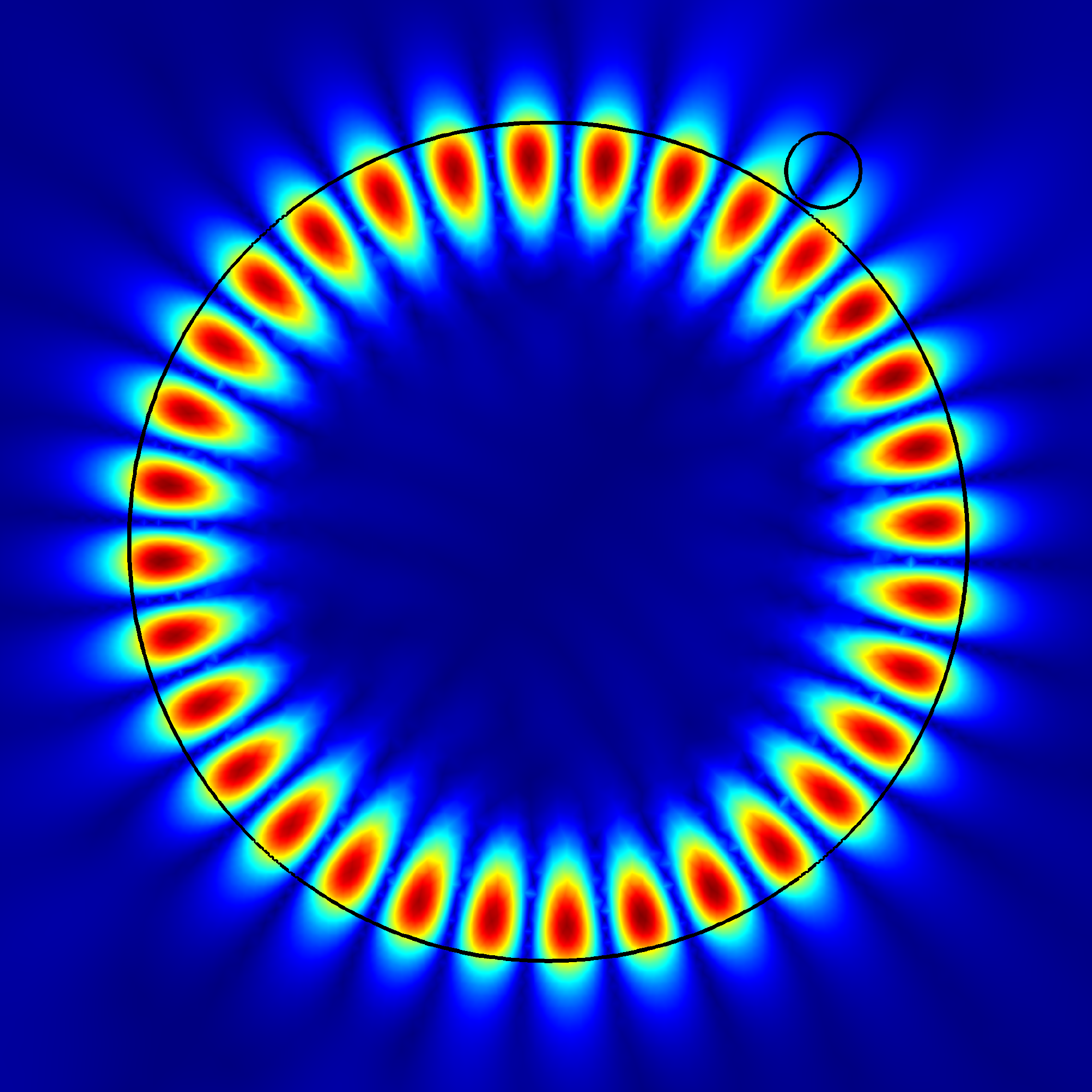}
 \includegraphics[width=0.47\columnwidth]{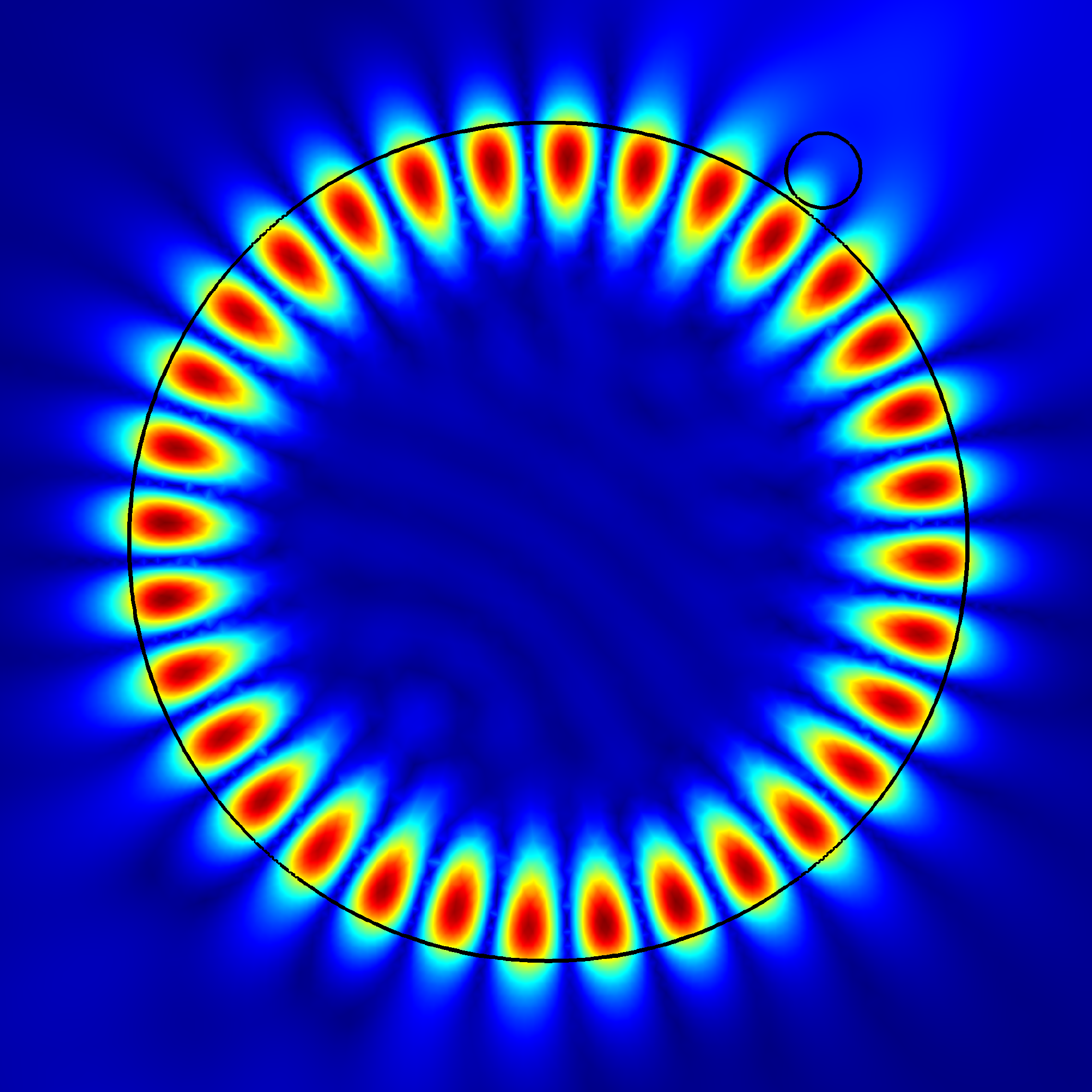}
\caption{
\label{fig:wgm_pair}
Calculated electric-field intensity distribution for the almost-degenerate WGM mode pair in the geometry of Fig.~\ref{fig:sweeps}, for $r/R = 0.089$, $d/R = 0.013$ and $\beta = 0.933$ [corresponding to $m=16$, $n=9$ in Eq.~\eqref{eq:betacond}].
Both modes have similar frequencies but different lifetimes: $\Omega_s = 9.87973 - 0.00088\,i$ (left) and $\Omega_c = 9.87977 - 0.00495\,i$ (right). This can be explained by the fact that the perturbation is broad enough to overlap with the field distribution of both modes, changing the frequency of both by a similar amount, but induces more scattering to the outside for the mode on the right, whose mode profile is symmetric  with respect to the nanoparticle position.}
\end{figure}

This perturbative perspective is useful due to several reasons. Firstly, it concretely connects the features of realistic microresonators with the effective two-mode Hamiltonians such as given in Eqs. \eqref{eqn:a},  \eqref{eq:hsw} and \eqref{eq:hwgm}. Secondly,
it shifts the problem from finding a resonator with the appropriate effective Hamiltonian to the problem of finding an adequate local perturbation, which has already been studied extensively \cite{Rubin.2010, Deych.2011}. Finally, the resulting conditions are very general and apply to all perturbations that only detune the imaginary parts of the eigenfrequencies.

We focus on a particularly versatile geometry, that of a circular resonator perturbed by a nanoparticle [see Fig.~\ref{fig:sweeps}(a)], which has been studied extensively both theoretically \cite{Wiersig.2011b} as well as in the context of sensing applications \cite{Foreman.2015,Chen.2017}.
The resonator is modelled as a circular dielectric disk of radius $R$ and representative refractive index $n=2$. The nanoparticle is of the same refractive index, has a radius $r$ and is placed at a distance $d$ from the disk. For the following discussion it is useful to introduce the dimensionless values $r/R$ and $d/R$ as well as the dimensionless frequency $\Omega = \omega R/c$.

In order to find a value where the relative shift $\delta$ is purely imaginary, we performed a parameter sweep for different values of $d/R$ and $r/R$ using full numerical simulations of the individual resonators with the methods described in Sec. \ref{sec:sys}, and compared the resulting eigenfrequencies with the unperturbed system, see Fig.~\ref{fig:sweeps} (b,c).
Based on these results, we find that we generically can identify suitable conditions by tuning the single parameter $r/R$ at fixed $d/R$.
For example, for the azimuthal mode number $m = 16$ and fixed $\beta=19 \pi/64\approx 0.933$, $d/R = 0.013$,  this occurs for $r/R =0.089$.  These conditions meet the constraint \eqref{eq:betacond}, where we chose $n=9$ to achieve a situation where the perturbing nanoparticle is located far away from the inter-resonator coupling regions in the chain configuration.
The field distributions of the two eigenmodes for this parameter combination are shown in Fig.~\ref{fig:wgm_pair}.

\subsection{System of coupled resonators\label{sec:chain}}

\begin{figure}[t]
\includegraphics[width=0.975\columnwidth]{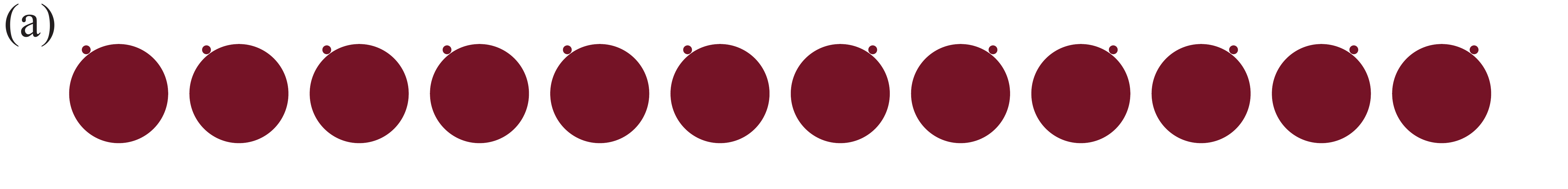}
\includegraphics[width=\columnwidth]{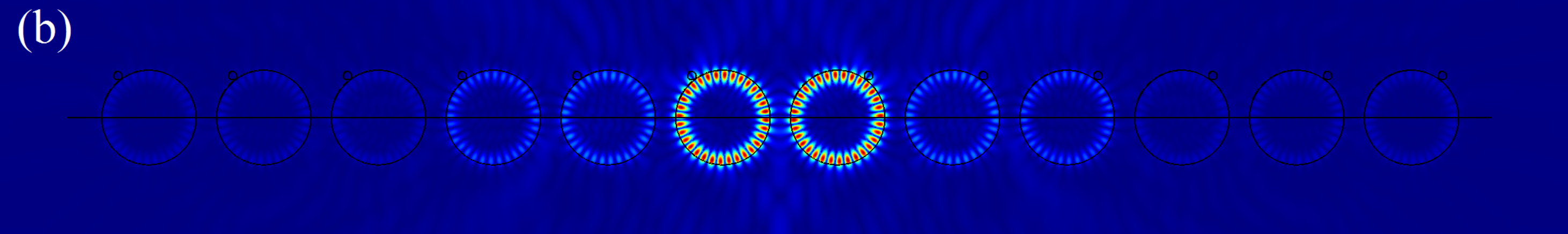}
\includegraphics[width=\columnwidth]{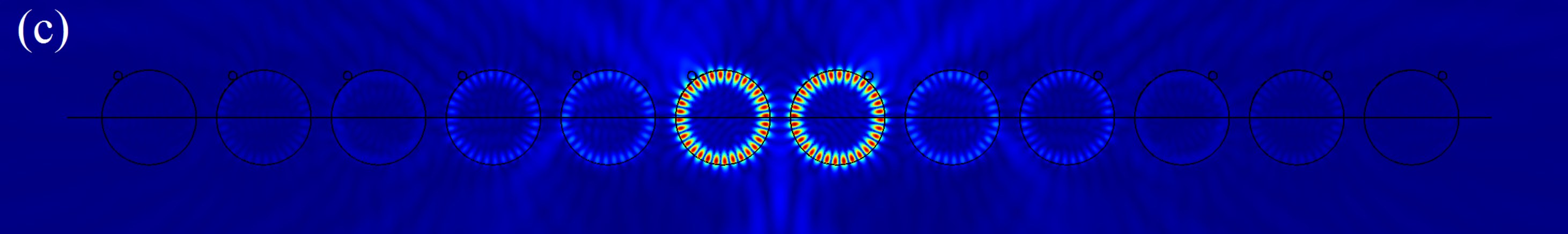}
\includegraphics[width=\columnwidth]{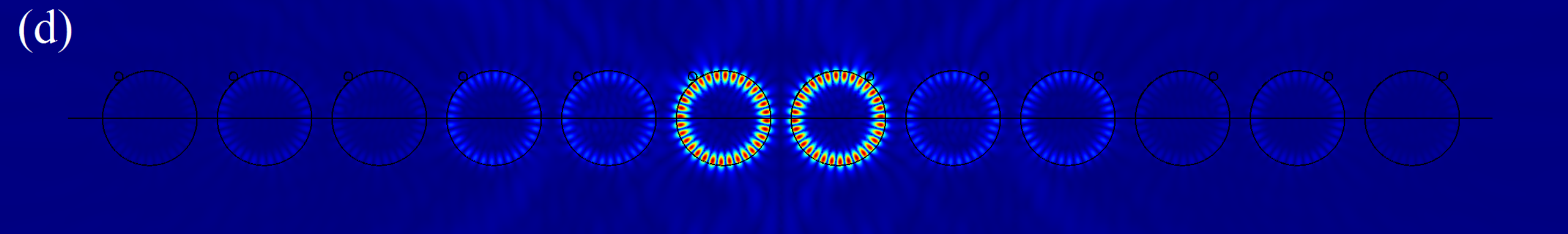}
\includegraphics[width=\columnwidth]{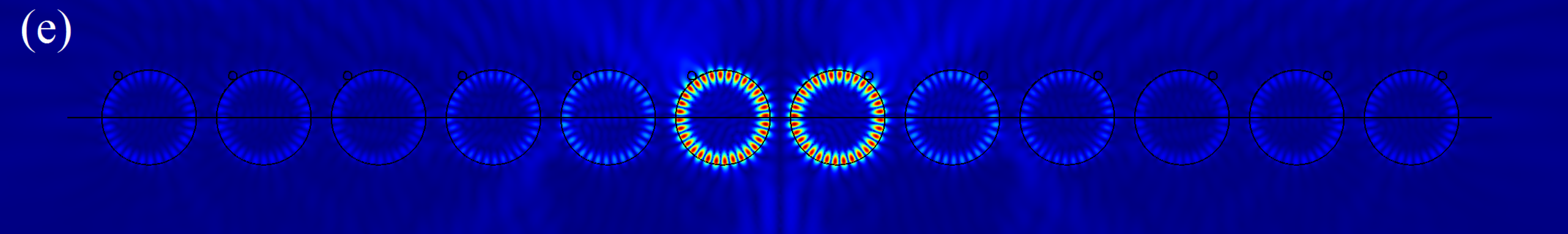}
\includegraphics[width=\columnwidth]{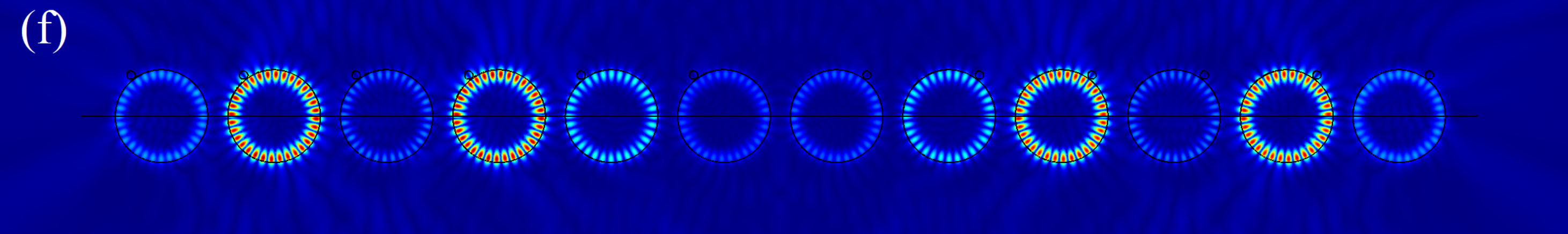}
\caption{(a)
Geometry of the designed CROW system, consisting of $N=12$ circular microresonators perturbed by nanoparticles that are placed at opposite positions on both sides of the central interface. The parameters for the individual parameters to the right of the interface are $r/R = 0.089$, $d/R = 0.013$ and $\beta_r = 0.933$, corresponding to those used in Fig.~\ref{fig:wgm_pair}, while to the left of the interface $\beta_l = \pi - \beta_r$. The inter-resonator spacing is set to $a/R = 0.43$, resulting in a sufficiently weak coupling so that the effective nonhermiticity is large.
(b-f)
Electric-field intensity distribution for selected eigenmodes. The first four states (b-e) correspond to the desired quadruplet of defect states, and can be interpreted as bonding and antibonding combinations that furthermore either have a minimum or a maximum at the position of the perturbing nanoparticles. The bottom panel (f) contrasts these with a representative extended state, for which the intensity is not localized around the central resonators.}
\label{fig:fielddistribution}
\end{figure}

\begin{figure}[t]
\includegraphics{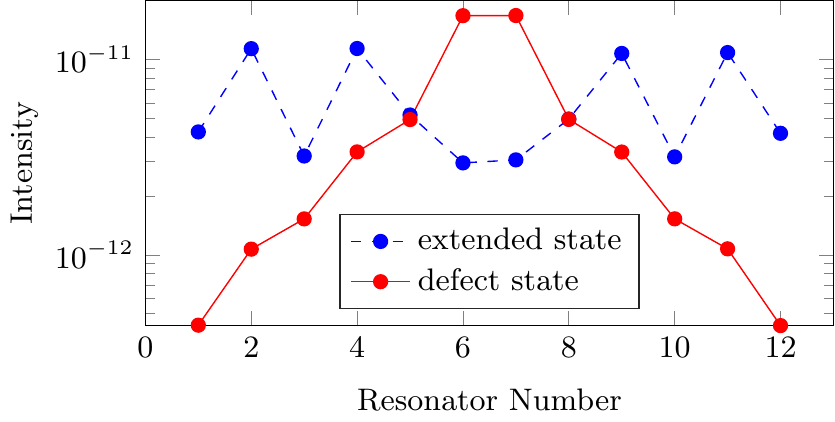}
\caption{
Electric field intensity in each resonator for the defect state in panel (c) and the extended state in panel (f) of Fig.~\ref{fig:fielddistribution}. The connecting lines are used only as a visual aid.
The defect state is characterized by an exponential decay away from the interface, as expected analytically for infinite chains \cite{Malzard.2015, Malzard.2018}. The slight deviations can be linked to the finite size of the chain. }
\label{fig:LineDistribution}
\end{figure}

\begin{figure}[t]
\includegraphics{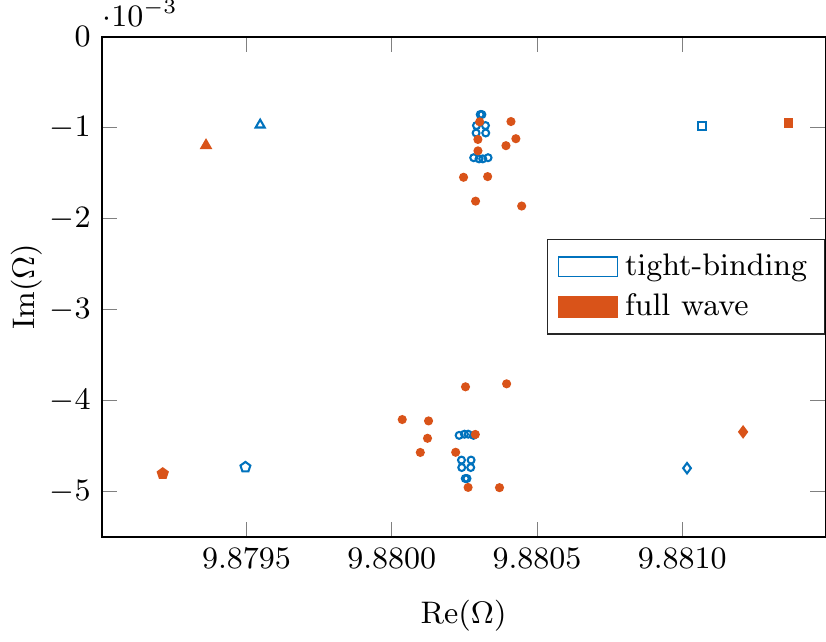}
\caption{
Complex resonance eigenfrequencies of the resonator chain shown in Fig.~\ref{fig:fielddistribution},
where we compare the result of the full wave calculations (red) to those of the tight-binding approximation (blue). In the tight-binding model, $\delta = 0.00203\,i$ and $W=0.00076$, resulting in the effective parameters $A/W=-B/W=-2.67$ (given that $n$ in Eq.~\eqref{eq:ab1} is odd).
The full wave solution confirms the expected spectral isolation for the defect states (non-circular symbols), as well as a good resemblance of a fourfold symmetry. The discrepancy with the results from the tight-binding model (where the defect states are denoted by corresponding unfilled symbols)
can be attributed to the two-mode approximation.
}\label{fig:EF_map}
\end{figure}

Whether the obtained lifetime difference is enough to induce the desired defect states should now depend on how it compares with the inter-resonator coupling strength $W$. We determined
this value by comparing the eigenfrequencies of the isolated resonator to the eigenfrequencies of a dimer of two coupled resonators for different inter-resonator distances $a$. This delivers an essentially exponential dependence of $W$ on $a$, as expected from the evanescent-mode nature of the coupling, which therefore can be adjusted easily over a large range of values by selecting a convenient resonator spacing.

Based on these preparations, we finally turn to the coupled-resonator geometry with the interface, which corresponds to a chain of circular resonators decorated by nanoparticles at position $\beta$ to the right of the interface, and position   $\pi-\beta$ to the left of the interface. We use the parameters  $\beta=19 \pi/64\approx 0.933$, $r/R =0.089$, and $d/R = 0.013$ determined above, so that the nanoparticles remain placed far away from the coupling regions of the adjacent resonators. The resulting geometry of the system can be seen in the upper panel of Fig.~\ref{fig:fielddistribution}.

The remaining panels in Fig.~\ref{fig:fielddistribution} show representative eigenmodes obtained by solving the full wave equation by the method described in Sec. \ref{sec:sys}. The four modes corresponding to defect states are localized at the interface and decay exponentially, whereas the rest of the states have wavefunctions which extend over the whole system, as shown for one example in the bottom panel.

For a more quantitative view we plot in Fig.~\ref{fig:LineDistribution} the field intensity in each single resonator, where we compare the second defect state depicted in Fig.~\ref{fig:fielddistribution}(c) with the extended state in Fig.~\ref{fig:fielddistribution}(f). This clearly demonstrates the strong confinement of the defect state around the interface region.

To further analyze these results, we compare in Fig.~\ref{fig:EF_map} the numerically determined resonance eigenvalues of the full wave calculations with the corresponding result from the tight-binding approximation. In both approaches, the four defect states are clearly separated from the extended states, and have lifetimes competing with the most long-lived extended states (for the two defect states in the upper region of the complex plane) and the most short-lived extended states (for the defect states further down in the complex plane).
The remaining discrepancies between the results can be linked to the two-mode approximation.
With help of this close correspondence, we can classify the four states as either bonding or anti-bonding, as well as either displaying a maximal or minimal intensity in  the region of the nanoparticle perturbation.  The two long-living defect states are then revealed as the bonding and antibonding combinations of states with a small amplitude around the nanoparticle location.

\section{Conclusion\label{sec:con}}
In this work we have shown that defect states in open coupled resonator systems are not necessarily tied to asymmetric backscattering, but can be present in a system with lifetime differences between the resonator modes. This situation can be achieved by suitable perturbations of simple symmetric resonator geometries, which change the lifetime of the  modes while keeping their real frequencies aligned. This simple design concept promises to be easier to implement than finding resonators with a good quality factor and asymmetric backscattering. More generally, this shows that genuinely nonhermitian defect states can be obtained by very simple means.
We implemented this concept for a system of coupled resonators perturbed by nanoparticles. The eigenfrequencies determined in full wave computations match well with the ones calculated in a tight-binding approximation focussing on a single whispering-gallery mode pair, realizing a quadruplet of defect states that have the expected localization at the interface. By their  perturbative nature, these results transfer to a wide range of different geometries and platforms, and thereby significantly broaden the scope of using defect states for photonic applications.

%

\end{document}